# Making life better one large system at a time: Challenges for UAI research

**Moises Goldszmidt**
Microsoft Research Silicon Valley
*moises@microsoft.com*

## Abstract

The rapid growth and diversity in service offerings and the ensuing complexity of information technology ecosystems present numerous management challenges (both operational and strategic). Instrumentation and measurement technology is, by and large, keeping pace with this development and growth. However, the algorithms, tools, and technology required to transform the data into relevant information for decision making are not. The claim in this paper (and the invited talk) is that the line of research conducted in Uncertainty in Artificial Intelligence is very well suited to address the challenges and close this gap. I will support this claim and discuss open problems using recent examples in diagnosis, model discovery, and policy optimization on three real life distributed systems.

## 1 Introduction

It is undeniable that services in the internet are changing the way we travel, access information, invest, bank, shop, and conduct business and research. These services are supported by an ecosystem of information technology (IT), including storage, network, and middleware/applications that is growing in complexity. The complexity is due to scale (over tens of thousands of computers in some cases), because systems are getting more distributed both in terms of function and geographical location, and because their ownership is becoming federated even inside an organization. These systems present numerous management challenges both in terms of day to day operations, and in terms of strategic and long term planning. Pervasive instrumentation and query capabilities are of course necessary elements for the solution to these problems [2, 23, 16]. In fact, there are now many commercial frameworks on the market for coordinated monitoring and control of large-systems: tools such as Hewlett-Packard's OpenView suite of products, IBM's Tivoli, and Microsoft's MOM, aggregate information from a variety of source and present it graphically to operators. Yet, it is widely recognized that the complexity and demands of these deployed systems surpasses the ability of humans to diagnose and respond to problems rapidly and correctly, and to assess strategic issues such as capacity planning [12, 7, 18]. Indeed, research on tools, algorithms, and technology for analyzing and interpreting the instrumentation data leading to (automation in) diagnosis, decision making, and control, has not kept pace with the demand for practical solutions in the field.

This paper (and the invited talk) makes the claim that UAI researchers are in a very advantageous position to address the challenges outlined above and advance the state of the art to close the gap between measurement and analysis for decision making. Section 2 enumerates the reasons for this claim and also outlines a couple of statistical inference problems that are particular to this domain and impactful for UAI research.

Section 3 reviews examples in diagnosis, model discovery, and policy optimization on three real-life distributed systems (Subsections 3.1, 3.2, and 3.3 respectively). These examples are drawn from my own research experience and projects. These projects are at different levels of maturity in terms of results. Yet, all of them I believe, open difficult challenges for UAI research and are illustrative of the space. As the projects relate to my own experience, this paper and the talk should not be taken as a comprehensive review of recent research in self-managing systems, datamining of systems logs, or machine learning techniques applied to systems. For that the reader is encouraged to look at the proceedings of the appropriate venues. Examples are *"The ACM Symposium on Operating*

*Systems Principles" (SOSP)*,[1] *"The Usenix Symposium on Operating Systems Design and Implementation" (OSDI)*,[2] *"The IEEE International Conference on Autonomic Computing" (ICAC)*, and the various workshops including those on *Hot Topics*, and those relating machine learning and systems.

Section 4 provides some final thoughts.

## 2 UAIers to the rescue!

Let me start by enumerating the reasons why I believe that UAIers are perfect for the job:

1. Given the relative maturity of measurement and instrumentation tools, as well as data collection and logging mechanisms, there are large amounts of data.
2. There is uncertainty coming from noise in the measurements, unobservable or hidden variables, missing values, and other factors.
3. To be really effective, the models for any of the tasks (diagnosis, decision making etc.) need to combine knowledge engineering and information coming from pattern recognition, and statistical learning (a definite strength of Bayesian networks [15]). The source of knowledge comes from the area of distributed systems and from the specific design of the particular systems in operation. At the same time, because of the complexity mentioned above, there is a need for inducing information from data in order to complement and sometimes verify this knowledge.
4. Models should work equally well in interaction with humans as well as in automated fashion. Therefore models should be both interpretable and susceptible to audits.
5. Some of the fundamental tasks and building blocks, diagnosis, knowledge discovery and engineering, decision making are core research issues in UAI.

The UAI reader will undoubtedly spot the obvious challenges for inducing models and making decisions in a large scale IT system: large amounts of data, high dimensionality, etc. These challenges are there, and they will undoubtedly require ingenious computational methods for inference. However, I would like to draw the reader's attention to two issues that are necessary building blocks and to the characteristics of this particular domain that introduces new challenges. The first issue pertains to hypothesis testing and model selection,[3] and the second issue pertains to the diagnosis of the parameters and structures of the graphical models we intend to use in this domain.

As part of our focus as a community includes the induction of models from data, our paths cross and intersect with research and applications of statistical inference. As we will see in Section 3, we need for example, to determine whether quantities of interest are different from one another in a "significant" way. This will take the form of comparing classification error in model selection tasks (diagnosis in Subsection 3.1), determining whether two probability distributions are the same for model discovery (Subsection 3.2), and in determining whether an action has an effect in the policy optimization task in Subsection 3.3. Whether these tests and decisions are embedded and therefore being taken in an automatic fashion or being taken by humans, in order to be trusted, we need to provide guarantees on false positives (and detection rates) or in terms of log-odds.

In discussing the first issue, I will first concentrate on a hypothesis testing approach, and then look at a Bayesian approach. Please note that in spite of the philosophical and methodological differences, statisticians focused on Bayesian approaches still devote considerable energy and effort on hypothesis testing [5, 6]. As a brief review of "classical" hypothesis testing and to introduce some useful notation: Consider the case where we observe data $X$ coming from a distribution $f(x|\theta)$ and that we are interested in testing $H_0 : \theta = \theta_0$ namely, the null hypothesis $H_0$ that the samples where generated from a distribution with parameter $\theta_0$. The statistical procedure consists of first choosing a test statistic $T = t(X)$ where large values of $T$ reflect evidence against $H_0$, and second computing the *p-value* $p = P_0(t(X) \geq t(x))$, rejecting $H_0$ if $p$ is "small enough". The objective is to test whether the p-value warrants the rejection of the null hypothesis; note that if the p-value does not reach the desired threshold all we can say is that we "failed to reject" the null hypothesis, but cannot say anything formal about its "true" state.[4] The statistics literature is full of examples of statistics $T$ and ways to compute p-values for most of the common cases. There are also variants depending on whether we want to test equality to $\theta_0$, whether $\theta \geq \theta_0$, or whether $\theta$ lies in some interval close to $\theta_0$. We won't go into these variants. The important thing for this paper is that there are two characteristics of the domain of application that makes the task particularly challenging: First, we must conduct a multiple number of those tests, and second, the number of samples used in the tests varies greatly from tens to hundreds of thousands.

[1] `www.sosp.org`.

[2] `www.usenix.org`.

[3] Depending on your leaning towards frequentist or Bayesian methodology.

[4] This is really a simplified version of hypothesis testing which suffices for the purposes of this paper. The reader is encouraged to go to the relevant literature for a formal treatment of this important subject (a good introduction can be found in [24]).

The p-value in hypothesis testing provides a basis to determine the false positive rate of a single test. The problem when we are conducting $m$ hypothesis testing is that the p-value determines the proportion of the null hypothesis falsely rejected on the $m$ tests. Yet, what we really want to know is the number of falsely rejected hypothesis amongst just the rejected tests. Thus for example, in subsection 3.2 rejecting the null hypothesis will mean that two services are correlated in a single computer. Those are the cases of interest for us. If in a server there are 10,000 tests, and we use a p-value of 0.05 as a basis for rejection, this would mean that we can expect the false positive rate to be 500. Now, if only 1,000 of those tests reject the null hypothesis, this would mean that half of those cases are not correlated whereas we think they are!

This issue has not escaped statisticians; the False Discovery Rate (FDR) approach [3] provides an algorithm for selecting the appropriate p-value amongst those obtained in the multiple tests to guarantee a given FDR $\alpha$ (i.e. an acceptable number of falsely rejected $H_0$ in the number of rejected tests). A search in any of your favorite engines for FDR will yield a large number of hits with *very* recent dates (ranging from 2000 to 2007) evidence of the importance of this issue in statistical research. In fact in a recent talk by Brad Efron [11], FDR is placed as a must directive for statistical research in the $21^{st}$ century.[5] He presents studies in particle physics and studies with microarrays in modern bioscience as his motivating domains. Yet, the main characteristics of these domains are very similar to the ones outlined above. I do think that UAIers should do well and pick up these techniques and adapt them to our models and techniques (and I am glad to realize that this has already started in a paper in this conference [20]).[6]

One of the problems with large sample sizes is that we not only have to worry about whether the differences are statistically significant (i.e., the test is rejected), but we should also check whether the size of the effect is too small. As an example, suppose we are trying to decide whether the difference between two means is significant (and to make it simple assume that we know that the distributions have the same variance). The usual statistic $T$ for the test in this case is the number of standard deviations that the sample mean is away from the mean in $\theta_0$. Note that as the number of samples increases, the value of the standard deviation decreases. This can happen to the point, where even if the test is rejected, the confidence interval is so close to zero, that the difference doesn't really make practical sense. This problem has also been addressed by statisticians in various ways, and has a longer history. One of the consequences is that we may have to select a "more appropriate" p-value (i.e., increase the number of standard deviations) to make sure that the test carries practical significance. Of course, this provides little comfort in our domain, since as stated above, there is great variability in the number of samples.

Berger and colleagues [4, 6, 5] make other technical points clear, including an attempt at calibrating the p-values when there are large number of samples and the difference between hypothesis testing, the risks of confusing a p-value with the posterior $P(H_0|X)$, and on ways of computing that posterior. This is a good transition to another approach consisting of following a strict Bayesian methodology and transforming the hypothesis testing into a log-odds test: Letting $M_0$ and $M_1$ be models for $H_0$ and $H_1$, in this case we compare the (log) ratio between $P(X|H_0, M_0, \theta_0)$ and $P(X|H_1, M_1) = \int P(X|H_1, M_1, \theta)P(\theta|M_1)$. The Bayesian approach is very attractive because it offers semantic clarity, it computes a posterior directly based on the observed data, and does not, in principle suffer from the multiple hypothesis testing problem. Yet, it does present us with some additional challenges: We now need to specify a model $M_1$ for $H_1$, namely the point of comparison to $H_0$, a parameterization $\theta$, and priors for $\theta$. This may not be trivial in some cases, and it may even be restrictive in others. In addition, we may find ourselves resorting to MCMC simulations to compute inferences (for example the integral above) which may be prohibitive for embedded tasks in our domain. Furthermore, we would also need a *threshold* for our log-odds decision and as exposed in [5] for example, this again may difficult and of course, application dependant. One is tempted to speculate that a suitable synthesis should emerge. Interestingly enough, Efron [11] also mentions such a synthesis, and makes a and explicit and strong connection to empirical Bayes methods in his own approach to the FDR problem.

The second major issue is that of diagnosis of graphical models themselves. It has been approximately 15 years since Cooper and Herskovits paper [10] on inducing Bayesian networks from data. Yet, in spite of valiant efforts for example [14, 21, 13, 20], we still don't have a set of tests, methodology, and software to test and diagnose over parameters and assumptions, recipes for remedial actions etc. We haven't agreed (or for that matter even discussed in depth) how to compute the "confidence" on the presence/absence of an arc or, on viable sensitivity tests for the appropriateness of a specific parameterization on a family in an induced model. This is particularly important in the domain of complex IT systems, where conditions may change drastically because of changes in workloads, daily regimes, and system configuration, and where our models that are automatically induced (see for example Subsection 3.1) are fundamental in enabling decision making. Basic and fundamental research on methodology for assessing the quality of induced models and remedial actions is badly needed.

[5] The paper is based on his presidential address to the American Statistical Association.

[6] Many thanks to D. Heckerman for sharing a preprint of this paper. I am not aware of other publications in UAI on the subject, if there are, my apologies for the omission.

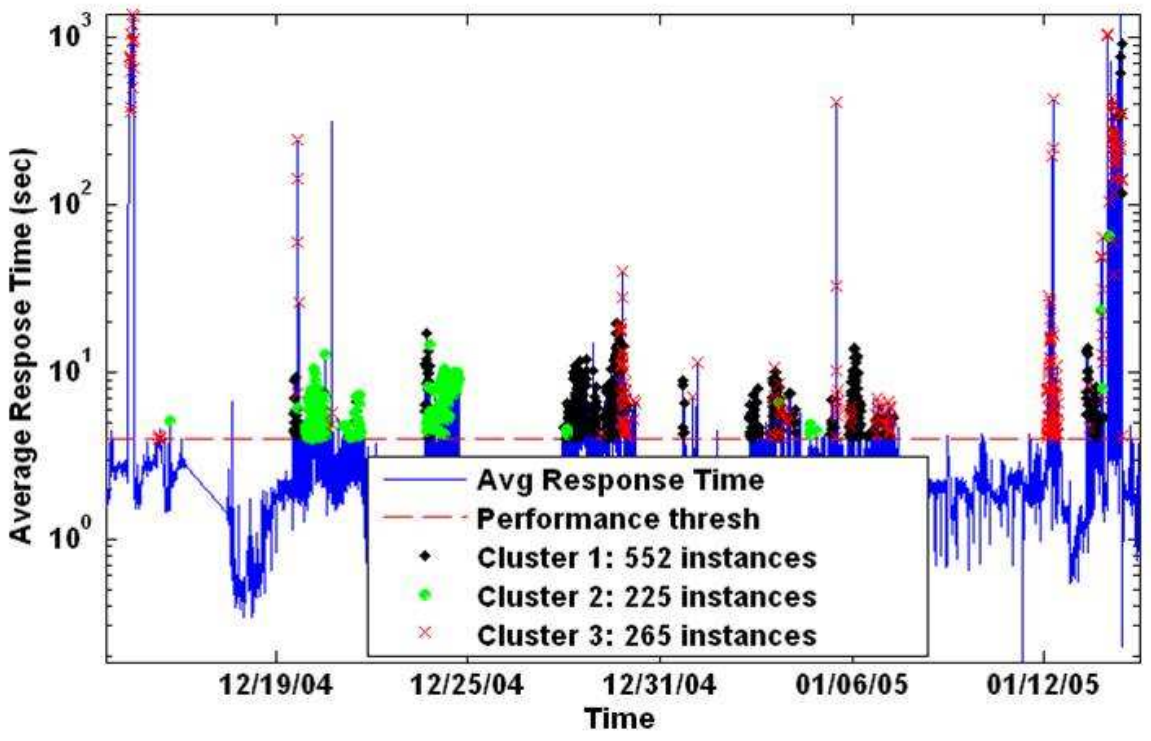

Figure 1: Time series of the average response time, annotated with instances of the different signatures for the metrics that present abnormal state. The signatures were clustered; cluster 3 refers to a recurrent problem identified by the system operators. Cluster 1 presents symptoms of high Database CPU utilization and low Application server utilization. Cluster 2 has memory and disk utilization metrics problems.

## 3 Three real systems and three tasks

As explained in the introduction, the following three subsections briefly describe three projects on real systems. These should be taken as descriptive and illustrative examples and not as set of prescriptive solutions and methodologies.

### 3.1 Diagnosis of performance problems: SLO compliance

Transactional systems are very common architectures in web e-commerce and in distributed database applications (such as desktop help or entitlement services). A common way to monitor the "health" of these systems is by checking the *Average Response Time* (ART) of the transactions. There are many reasons for this: a) it is a relatively easy metric to monitor given modern servers (just log the entry and exit times of the transaction, b) it is relatively easy to relate this time to some business objective of interest, and c) it is again a relatively easy metric to agree upon when different parts of the system belong to different organizations inside the same company (the front and middle tiers may belong to the IT department and the database may belong to a business division). Usually, the ART is part of a Service Level Objective (SLO) which in turn forms part of the contract between the different departments offering and receiving the service (sometimes even in the same organization). What is important to us is that a threshold $t$ on the ART is agreed upon whereby if the ART $> t$, then the SLO is said to be violated (sometimes these violations are connected to penalties under contracts). It is important to be able to diagnose these violations in order to assign responsibilities, and to identify performance problems, misconfigurations, and faults. The complete solution to this problem is still open.

In [8, 9] we provided one possible building block towards this solution. Operators in one transactional system were also collecting a set of low level metrics per each server and machine instance pertaining to cpu, disk, and memory utilization, processes counters etc. There was a total of approximately 30 metrics per server. Operators wanted help in trying to determine which metrics were correlated to each particular instance of an SLO violation. By inducing models correlating SLO violations with the statistics of the metrics, we were able to induce a *signature* of each violation in terms of the metrics. Using these signatures we not only were able to highlight which metrics were in abnormal states, but we enabled the identification of recurrent SLO violations, and by using these signatures as indexes we enabled the cataloging of the different types of SLO violations for search and retrieval of successful repair actions. Once the signature was induced these tasks only required well known clustering and retrieval algorithms. Figure 1 shows an output of the graphical interface of the analysis system we built, which included a "map" of the different clustered signatures overlaid on the time series of the ART. It took 80 pages of email and three weeks for every competent operators to identify all the instances in cluster 1. Our procedures did it in a matter of minutes when used as a forensic tool.

The approach we followed was that of inducing a Bayesian network classifier that used the low level metrics as features and the state of the SLO as a class variable. The advantage of using a classifier is that we had a very convincing metric to judge whether the model was indeed capturing the relationship between the low level metrics and the SLO violations: classification accuracy. Also, the use of a classifier, as opposed to a regression model from the metrics to the actual ART, enabled us to use relatively simple (and robust) parameterizations for the models. In order to build the signatures, we use Bayes rule and computed the log-likelihood of each metric (given its parents and the class). If the metric contributed to the classification of the SLO in the *violated* state, then the metric was deemed abnormal. The reader is directed to [8, 9] for details.

In order to increase the accuracy of the models to the low to mid 90's we had to resort to the use of ensembles of classifiers, feature selection, and updates on the parameters to make sure that the classifiers were able to deal with the workload changes on the system. Feature selection implied a search over the space of models where it was really important to compare two models and determine whether the difference in classification accuracy was significant. The reason was that signatures may change drastically and appear different even if the underlying cause would be identical. Equally important was the establishment of a sound criteria to determine and select the models from the ensemble that were applicable during different time periods. These are of course realted to the two main issues discussed in Section 2, and the definite answers are still open.

### 3.2 Model discovery: dependencies between computers and network services

Facilities such as remote file sharing and email invariably rely and depend on multiple network services ranging from directory services to authentication. These services are identified (to a first order of approximation) by the protocols used, e.g. SMB, DNS, HTTP, LDAP etc. Regular management functions such as determining provenance, planning, detecting change, and diagnosis rely on knowledge of the dependencies between the hosts comprising the network and the network services. As the system grows and evolves, it is difficult to keep track and validate these dependencies. Thus the automatic discovery and verification of these dependencies is an important unsolved problem.

In [1] we recently approached the problem of automatically discovering and maintaining a directed graph of these dependencies, where a node will represent a computer in the network (server, desktop etc.) and an arc will represent a protocol/service dependency. Figure 2 depicts an example of a fragment of such a graph. The data used to build that graph is from real network data obtained at Microsoft comprising headers and partial payloads of around 13 billion packets. The traces cover almost all the traffic transmitted and received by around 500 computers on the local network.

The approach in [1] consists of two phases. The first phase discovers dependencies locally on each computer/server and then a distributed algorithm builds the graph using these local dependencies. The graph(s) can be maintained centrally or locally on each computer/server. The approach is designed to be distributed, online, and therefore was constrained to only require the inspection of the packets headers (not the payload or content). The majority of issues of interest pertain to the local test.

The first test we devised was based on comparing two cumulative distributions using a variant of the Kolmogorov-Smirnov test [19]. The main idea is as follows. We abstract the input and output packets grouping them by protocol/service and source/destination id. Each one of these groupings is called a channel. We want to find out whether a specific output channel $B$ (say dns and destination $y$) depends on input channel $A$ (say http and source $x$). To do this we fit a Cumulative Distribution Function (CDF) of the time between arrival and departure of the input/output, and then compare it to the CDF of the time between the same input channel and a virtual output channel $R$ whose departure times are uniform and random. If the two CDFs are different then we declare the channels to be dependent (their behavior differs from that between the input and a random output). The initial results are very promising and certainly enough to improve the state of the art for the network management tasks outlined above. Yet, the issues for hypothesis testing outlined in Section 2 are present and need to be dealt with. First, each server has on the order of hundreds inputs and hundreds of outputs. Therefore we must perform multiple hypothesis tests which leads to the necessity for some kind of FDR. In addition, the DNS server contains channels with tens of samples, and also with hundreds of samples. The HTTP server contains channels with tens of thousands of samples.

We are currently systematically testing, validating and comparing our initial approach to a model selection approach (log-likelihood) based on Bayesian considerations.[7] In addition we are also testing and comparing FDR and an approach based on building more complex models between the inputs and outputs including Bayesian networks.

Finally, there are many issues with building and maintaining the service dependency graphs, exploring the changes over time, and learning their normal and abnormal behavior for diagnosis purposes.

[7]This Bayesian test was mainly designed by C. Bishop and J. MacCormick.

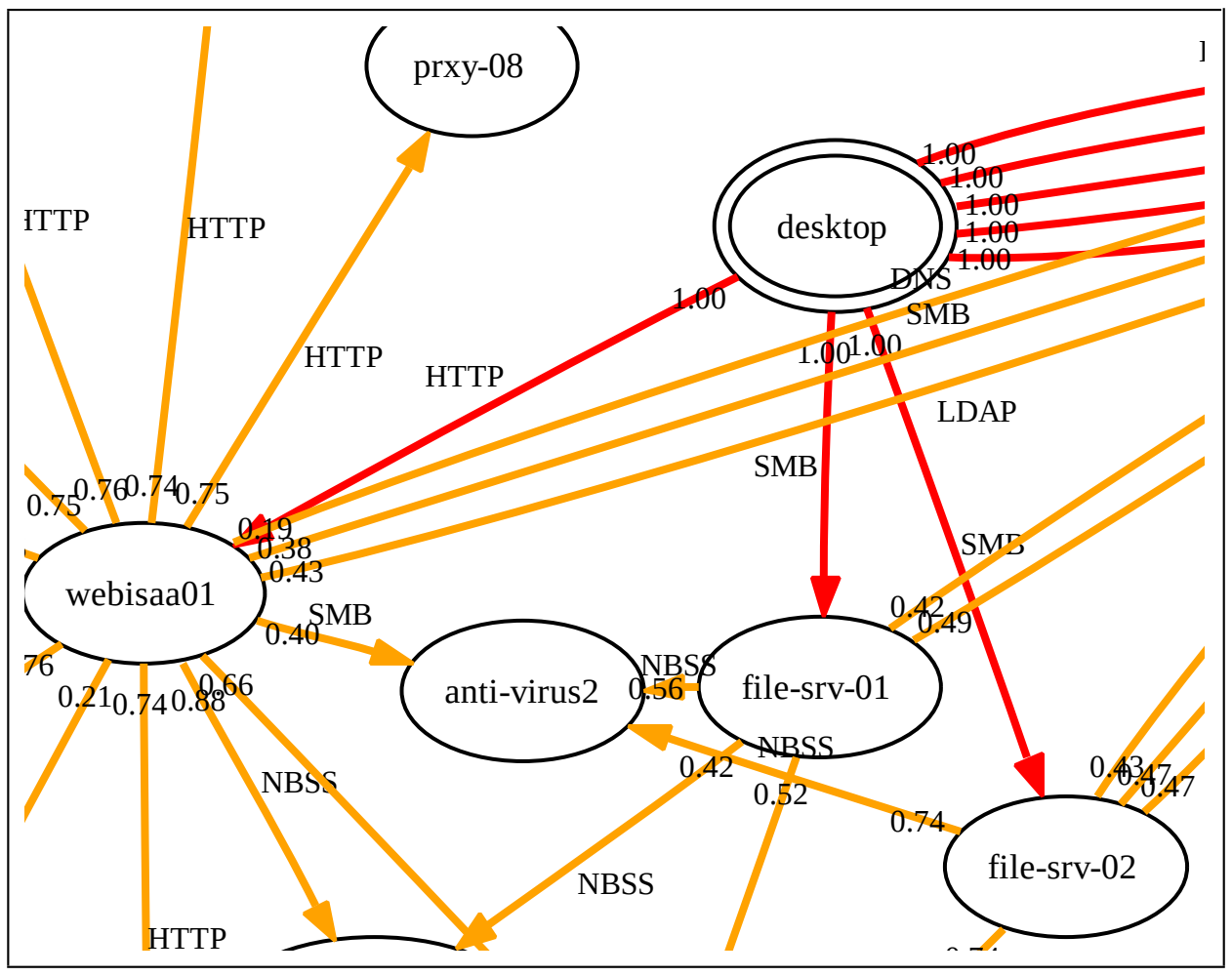

Figure 2: Small fragment of a constellation built automatically by Constellation depicting the dependencies of a single computer in an enterprise network. The machine "desktop" (doubly-circled in the figure) is the root of the constellation. The other nodes represent servers on which the root depends (either directly or indirectly), while the edges reflect the corresponding services that cause dependency.

### 3.3 Policy optimization: keeping the infrastructure running

Microsoft has developed an in-house infrastructure for automatic data center management called *Autopilot* [17]. Its design was primarily motivated by the need to keep the total cost of a data center, including operational and capital expenses as low as possible. The first version of Autopilot concentrates on the basic services needed to keep a data center operational: provisioning and deployment; monitoring; and the hardware lifecycle including repair and replacement. In this section we concentrate on the automatic repair services of the hardware lifecycle. Autopilot supports a simple model for fault detection and recovery. It was designed to be as minimal as possible while still keeping a cluster operational. The unit of failure is defined to be a computer or network switch rather than a process, and the only available actions for repair are *Reboot*, *ReImage*, *Replace*, and *DoNothing*. Faults are detected using a set of "(active) sensors" called *watchdogs*. A watchdog probes one or more computers to test some attribute, and then reports to a part of the system called the *Device Manager*. The Device Manager computes an error predicate for any computer using the watchdog attributes: if any watchdog reports *Error*, the computer is in error; if all watchdogs report either *OK* or *Warning* the computer is not in error. Once a computer is held to be in error, it may be unavailable for a substantial period. A computer that is functioning normally is marked by the Device Manager as being in the *Healthy* state. If any watchdog reports an error for that computer it is placed in the *Failure* state and assigned an appropriate "recovery action" from the set above. The choice of recovery action depends on the recent repair history of the computer and the error that is reported. We call this choice a policy.

There is a logging service that records watchdog reports, Device Manager assigned state, and repair action. There are many important questions that can be potentially answered by mining these logs: How reliable are the different watchdogs, and can we estimate and then minimize the false positives in watchdogs? Can we capture transient errors? Can we capture correlated faults? Can we evaluate the effectiveness of the current policy and synthesize an optimal policy? Most of the procedures and algorithms to provide sound answers to these questions will require progress on the issues outlined in Section 2. Also, as these data centers contain tens of thousands of computers (and this number is growing), these algorithms and procedures need to scale in order to deal with the volume.

Finally, it is enticing to think that as these logs contain the set of actions taken, plus evidence of their effects, it may provide a fertile ground for testing algorithms dealing with causal discovery and modeling [22].

## 4 Final Remarks

The examples I present above clearly illustrate that UAI based techniques, methods, and algorithms, can be useful in automatically discovering correlations and dependencies in complex networked computer systems, that ultimately lead to better models and optimal decision making. The expected consequence of this is that the services we have come to rely upon will work optimally and uninterrupted.[8] At the same time, I trust that the exposition above makes it clear that this domain also contains a set of challenges that will definitely advance the state of the art in UAI. However, my objective with this paper and the talk was not to present a list of research topics or what I think are future "must do" for UAI research. Rather, my intention has been to describe a fertile ground for research and application of UAI based techniques hoping to entice the imagination of colleagues in the field.

[8] It has been my experience in retreats and systems oriented forums that there is sometimes a myopic held belief that the only usefulness of "statistical learning theory" (as sometimes machine learning, belief network, and statistical induction research is grouped) is in automatically detecting abnormalities.

## Acknowledgements

Many thanks to research colleagues and collaborators in [1, 8, 9]. In particular to I. Cohen and J. MacCormick for numerous discussions on these topics. Thanks to M. Isard for discussions on Autopilot and for the material in Section 3.3. K. Achan has spent a lot of time with me mining Autopilot logs.